\documentclass[11pt,a4paper]{article}
\usepackage{amsmath,amssymb}
\usepackage{epsfig,graphicx}
\begin{document}

\title{\bf ENERGY LOSS IN STOCHASTIC ABELIAN MEDIUM}
\author{M.~R.~Kirakosyan$^a$\footnote{{\bf e-mail}: martin.kirakosyan@cer.ch; supported by the RFBR grant 08-02-91000},
A.~V.~Leonidov$^{a,b}$\footnote{{\bf e-mail}: leonidov@lpi.ru; supported by the RFBR grants 06-02-17051 and
08-02-91000}}
 \date{}
 \maketitle
\begin{center}
 $^a$ \small{\em Theoretical Physics Department, P.N.~Lebedev Physics Institute}\\
 \small{\em Leninsky pr. 53, Moscow, Russia}\\
 $^b$ \small{\em Institute of Theoretical and Experimental Physics}\\
 \small{\em Bolshaya Cheremushkinskaya 25, Moscow, Russia}
\end{center}

\begin{abstract}
Energy losses of fast charged particles in randomly spatially inhomogeneous stationary medium are considered.
Analytical results for effective dielectric permittivity are presented.
\end{abstract}

Studies of energy losses of fast particles in a medium provide important information on it properties. Of special
interest are situations in which the studied medium is not homogeneneous. A particular case that draws a lot of
attention is that of a medium that is homogeneous only on average and characterized by random inhomogeneities on
the event-by-event basis. This situation is common in radiophysics and acoustics, see e.g. \cite{BKRT70}. In
\cite{GRZ97} it was shown that the parton system created at early stages of ultrarelativistic heavy ion collisions
is characterized by violent event-by-event fluctuations of local partonic energy-momentum density. Progress in
understanding the properties of energy loss of fast charged particles in such media generalizing the analysis
referring to the homogeneous case, see e.g. \cite{B82,GT92}, is therefore very important in understanding the
characteristics of dense matter created in ultrarelativistic heavy ion collisions. In the present talk we present
estimates of the energy loss of fast particle in the random inhomogeneous medium in the abelian approximation
based on the analytical calculation of the effective dielectric tensor. A detailed description of this calculation
and its generalization to the case of quark-gluon plasma will appear in the forthcoming publication \cite{KL08}.

It is well known that the energy losses of charged particles in the medium can be computed by the work done on the
particle by the electric field it creates in this medium. In the random medium the average losses per unit length
$dW/dz$ of a fast particle with charge $e$ and velocity ${\bf v}$ are thus determined by the average electric
field:
\begin{equation}\label{genloss}
 \frac{dW}{dz} = e \frac{\bf v}{v} \langle {\bf E} ({\bf r},t) \rangle_{{\bf r}={\bf v}t}
\end{equation}
In the case of inhomogeneous medium the local fluctuations of its properties are customarily parametrized by the
coordinate-dependent dielectric permittivity $\varepsilon({\bf r})$. For given configuration of $\varepsilon({\bf
r}$ the spectral component of the electric field is determined from
\begin{equation}\label{eqE1}
 \triangle {\bf E} - \nabla \left( \nabla {\bf E} \right) + \omega^2 \, \varepsilon ({\bf r}) = 4\pi i \omega j(\omega)
\end{equation}
where $j(\omega)$ is a spectral component of the external current which in the considered case of uniformly moving
fast particle reads  ${\bf j}(\omega,{\bf k})=2 \pi e {\bf v} \delta (\omega - {\bf k} {\bf v})$. Random
inhomogeneities can conveniently be parametrized by explicitly identifying the non-random and random contributions
to dielectric permittivity:
\begin{equation}
 \varepsilon({\bf r}) \, = \, \varepsilon_0 \left(1+\xi ( {\bf r} )  \right)
\end{equation}
where $\xi( {\bf r} )$ is a random contribution to permittivity having zero mean $\langle {\xi(\bf r}) \rangle=0$.
In what follows we shall consider the simplest case of Gaussian ensemble so that $\xi({\bf r})$ are fully
characterized by binary correlation function
\begin{equation}
 \langle \xi \left( {\bf r}_1 \right) \xi \left( {\bf r}_2 \right) \rangle=g\left( |{\bf r}_1-{\bf r}_2|\right)
\end{equation}
It is well known that taking into account the random inhomogeneities leads to an equation for the average electric
field containing a tensor of effective dielectric permittivity $\varepsilon_{ij}(\omega,{\bf k})$ that depends on
the statistical properties of fluctuations $\xi({\bf r})$:
\begin{equation}\label{eqE2}
\left [\varepsilon_{ij}(w, \mathbf{k})-\frac{k^2}{w^2}\left ( \delta_{i,j} - \frac{k_{i}k_{j}}{k^{2}} \right )
\right]E_j(w, \mathbf{k})=\frac{4 \pi}{i \, w}j_i(w, \mathbf{k})
\end{equation}
where we have introduced a notation $w=\sqrt{\varepsilon_0} \omega$. It is convenient to explicitly introduce the
transverse and longitudinal components $\varepsilon_t$ and $\varepsilon_l$:
\begin{equation}\label{epstl}
 \varepsilon_{ij} (w,{\bf k}) \equiv \left(\delta_{ij}-\frac{k_ik_j}{k^2} \right) \varepsilon^t (w,{\bf k}) +
 \frac{k_ik_j}{k^2} \varepsilon^l (w,{\bf k})
\end{equation}
In terms of the decomposition (\ref{epstl}) the expression for the energy losses takes, for some given
$\varepsilon_{ij}(\omega,{\bf k})$, the form
\begin{eqnarray}\label{losstl}
 \frac{dW}{dz} &=& -\frac{e^2}{2 \pi^2 v} \int d^3k
 \left\{ \frac{w}{k^2} \left[ {\rm Im} \frac{1}{\varepsilon^l(w,{\bf k})} \right. \right. \\
 \phantom{a} &-& \left. \left. \left( w^2-v^2k^2 \right) {\rm Im} \frac{1}{w^2 \varepsilon^t(w,{\bf k}) -k^2} \right]
 \right\}_{w={\bf k}{\bf v}} \nonumber
\end{eqnarray}

Let us now consider the particular case of an exponential binary correlation function
\begin{equation}\label{corfun}
 g(r)=\sigma^2 e^{-r/a}
\end{equation}
that allows an explicit analytical computation of $\varepsilon_{ij}(\omega,{\bf k})$ in the one-loop approximation
corresponding to the regime $\sigma^2 (wa) \ll 1$ \cite{KL08}. The corresponding calculation is naturally
performed in terms of the polarization tensor $\Pi_{ij}(w,{\bf k})$ related to dielectric permittivity by the
following relation:
\begin{equation}\label{epsPi}
 \varepsilon_{ij} (w,{\bf k}) = \varepsilon_0 \left(1-\frac{1}{w^2} \Pi_{ij}(w,{\bf k}) \right)
\end{equation}
The decomposition of dielectric permittivity (\ref{epstl}) leads to the corresponding decomposition of the
polarization tensor. Explicit expressions for its transverse and longitudinal components $\Pi^t_{ij} (w,{\bf k})$
and $\Pi^l_{ij} (w,{\bf k})$ read
\begin{eqnarray}\label{Pit}
 \Pi^t_{ij} (w,{\bf k})& = & \sigma^2 w^2
 \left[
    \frac{w^2}{(w+i \delta)^2-k^2}-\frac{\delta(\delta+iw)}{2k^2} \right.\\
   \phantom{a} & + & \left. \frac{\delta^2+w^2+k^2}{k^2} \frac{\delta}{k} \arctan \left( \frac{ik}{w+i\delta}\right) \nonumber
 \right]
\end{eqnarray}
and
\begin{eqnarray}\label{Pil}
 \Pi^l_{ij} (w,{\bf k})& = & \sigma^2 w^2
 \left[
    1+\frac{\delta(\delta+iw)}{2k^2} \right.\\
   \phantom{a} & - & \left. \frac{\delta^2+w^2+k^2}{k^2} \frac{\delta}{k} \arctan \left( \frac{ik}{w+i\delta}\right) \nonumber
 \right]\,,
\end{eqnarray}
where $\delta=1/a$. Let us stress that even if the non-random delectric permittivity $\varepsilon_0$ does not have
a significant imaginary part so that the corresponding energy losses determined by (\ref{losstl}) are absent, the
effective dielectric permittivity determined by (\ref{Pit},\ref{Pil}) has a nontrivial imaginary part and,
therefore, there arise specific energy losses directly related to the random fluctuations in the medium in which
they propagate.

Let us present a few numerical calculations of this stochastic energy loss. In presenting the results it turns out
convenient to rewrite the expression for the energy loss (\ref{losstl}) in the form:
\begin{equation}
 \frac{dW}{dz} = - \frac{e^2}{\pi} \frac{1}{a^2} \left[f_t(pa)+f_l(pa) \right]
\end{equation}
where $p$ is a momentum of the projectile\footnote{In numerical computations the integration over momentum of
field components is cut at the scale $p$.} Let us first consider the case $\varepsilon_0=0.7$ (i.e. without
Cherenkov radiation) and $\sigma=0.3$ and consider the stochastic energy loss as a function of the dimensionless
variable $pa$. The result, broken into separate transverse and longitudinal contributions $f_t(pa)$ and $f_l(pa)$,
is shown in Fig. 1. We see that in this sub-Cherenkov regime longitudinal losses dominate over transverse ones. It
is of interest to compare these results with the energy losses with the case in which Cherenkov losses exist
already for the non-random case. To this aim let us consider the case $\varepsilon_0=1.1$ and the same value of
$\sigma=0.3$. The results are shown in Fig. 2. We see that the transverse losses are significantly amplified in
comparison with the case of $\varepsilon_0=0.7$ and are dominant with respect to the longitudinal losses. It is
also important to understand a dependence of the energy losses on the fluctuation magnitude $\sigma$. This
dependence is shown, for $\varepsilon=0.7$ and the particular value of particle momentum $pa=1$, in Fig. 3. We see
that the dependence on $\sigma$ is quite pronounced and the losses rapidly grow with growing $\sigma$.

\begin{center}
{\bf Conclusions}
\end{center}

Let us formulate the main results of the present talk:
\begin{enumerate}
 \item{Energy losses of charged particle in randomly inhomogeneous medium based on an
 analytical calculation of the effective dielectric permittivity were considered.}
 \item{A significant difference between transverse and longitudinal components was demonstrated.}
 \item{A rapid growth of energy loss with growing fluctuation magnitude $\sigma$ was found.}
\end{enumerate}

\begin{figure}[ht]
\epsfig{file=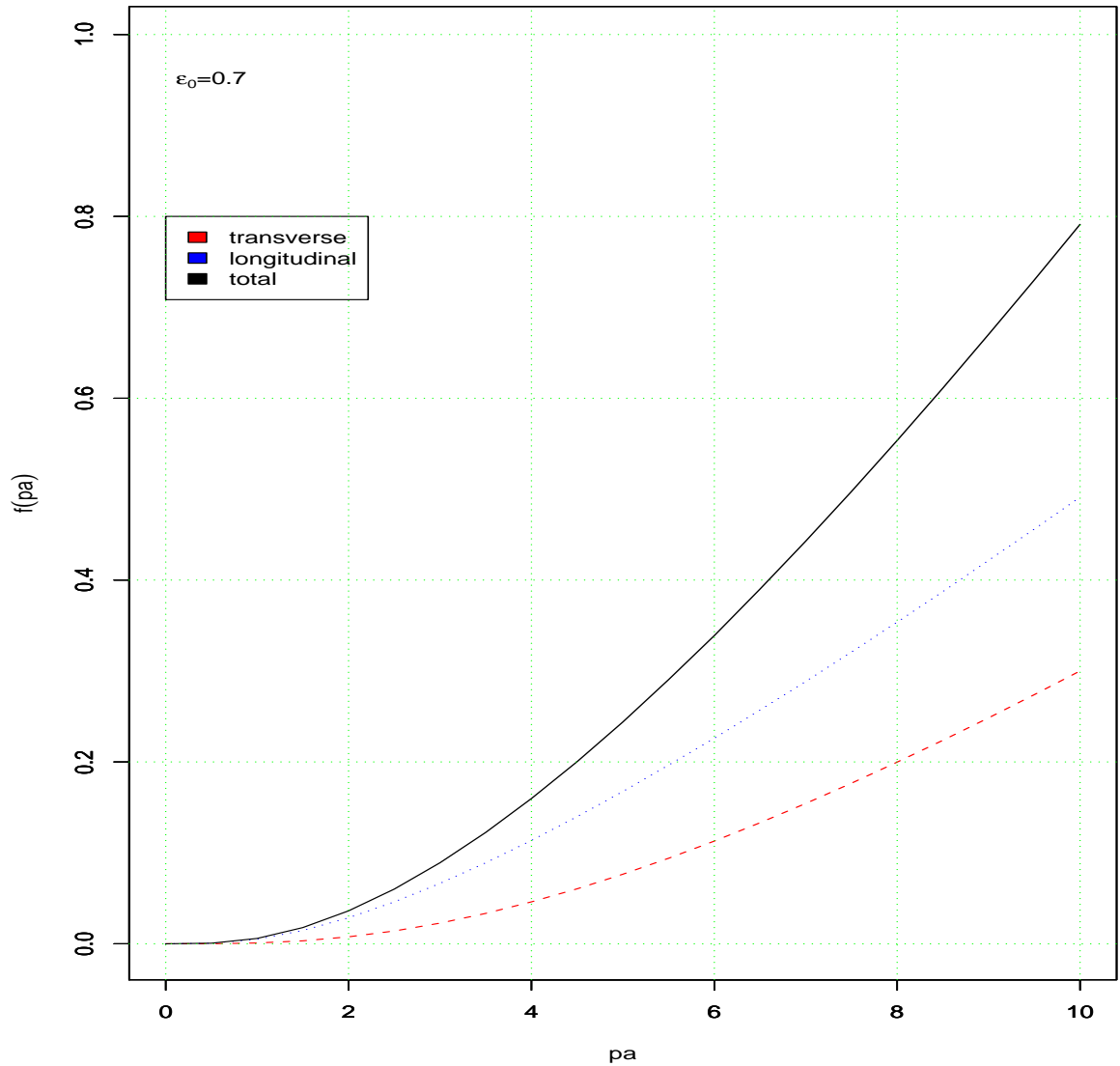,height=16cm,width=16cm}
 \caption{Stochastic energy losses for $\varepsilon_0=0.7$ as a function of momentum scale $pa$:
 transverse losses $f_t(pa)$- dashed line (red); longitudinal losses $f_l(pa)$ - dotted line (blue);
 total losses - solid line (black).}
\end{figure}

\begin{figure}[ht]
\epsfig{file=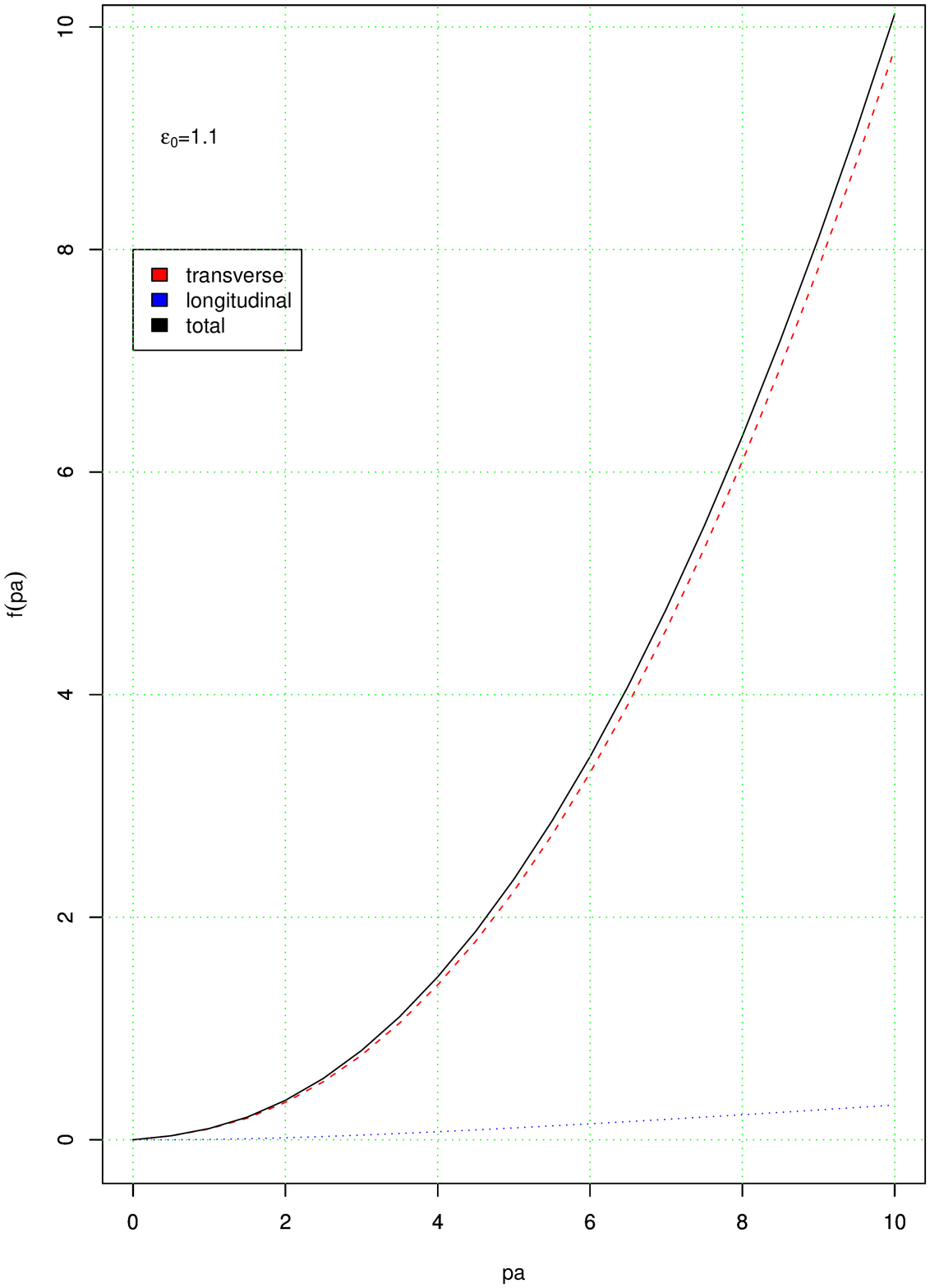,height=16cm,width=16cm}
 \caption{Stochastic energy losses for $\varepsilon_0=1.1$ as a function of momentum scale $pa$:
  transverse losses $f_t(pa)$- dashed line (red); longitudinal losses $f_l(pa)$ - dotted line (blue);
  total losses - solid line (black).}
\end{figure}

\begin{figure}[ht]
\epsfig{file=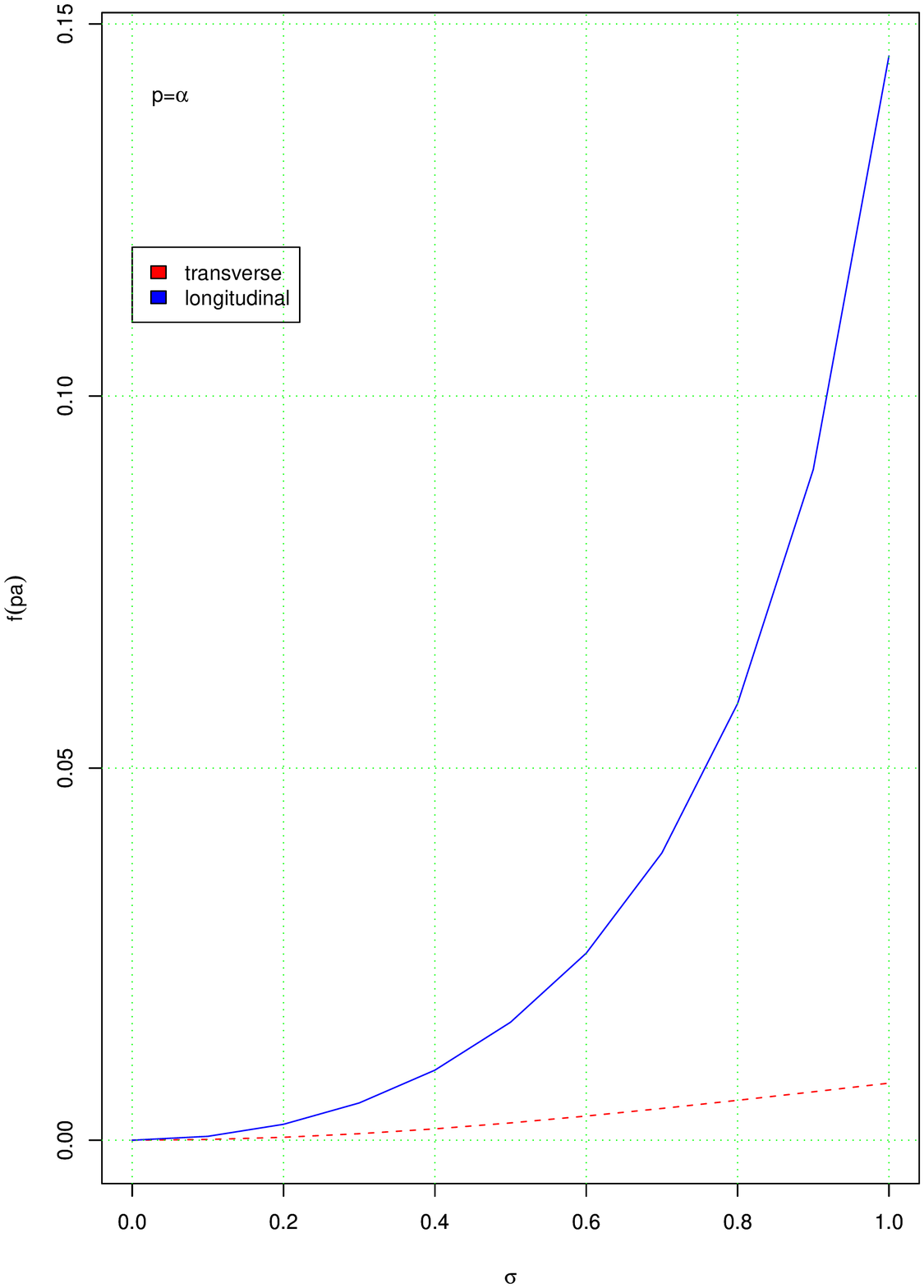,height=16cm,width=16cm}
 \caption{Dependence of stochastic energy losses on the fluctuation magnitude $\sigma$ at $p=1/a$: transverse
 losses - dashed line (red); longitudinal losses - solid line (blue).}
\end{figure}



\begin{thebibliography}{99}

\bibitem{BKRT70}
Yu.N. Barabanenkov, Yu.A. Kravtsov, S.M. Rytov, V.I. Tatarsky, {\it UFN}\ {\bf 102} (1970), 3 (in Russian)

\bibitem{GRZ97}
M. Gyulassy, D.H. Rischke, B. Zhang, {\it Nucl. Phys.}\ {\bf A613} (1997), 397

\bibitem{B82}
J.D. Bjorken, {\it Energy Loss of Energetic Partons in Quark-Gluon Plasma: Possible Extinction of High $p_T$ Jets
in Hadron-Hadron Collisions}, preprint FERMILAB-Pub-82/59-THY

\bibitem{GT92}
M. Gyulassy, M. Thoma, {\it Nucl.Phys.}\ {\bf A544} (1992), 573

\bibitem{KL08}
M. Kirakosyan, A. Leonidov, in preparation



\end{thebibliography}
\end{document}